\documentclass[preprint,aps,12pt,showpacs,nofootinbib,tightenlines]{revtex4}
\usepackage{amsmath}
\usepackage{amssymb}
\usepackage{epsfig}
\usepackage{graphicx}
\newcommand{\pvsl}{ p \hspace{-2.0truemm}/_{K^*} }
\newcommand{\psl}{ P \hspace{-2.4truemm}/ }

\newcommand{\esl}{ \epsilon \hspace{-2.1truemm}/ }
\def\be{\begin{eqnarray}}
\def\en{\end{eqnarray}}
\def\non{\nonumber\\}

\def\ra{\rangle}

\def\sl{\!\!\!\slash}
\def\prd{{Phys. Rev. D}~}
\def\prl{{ Phys. Rev. Lett.}~}
\def\plb{{ Phys. Lett. B}~}

\def\epjc{{ Eur. Phys. J. C}~}
\newcommand{\acp}{{\cal A}_{CP}}
\begin{document}
%%--------------------------------------------
\title{Study of scalar meson $f_0(980)$ from $B \to f_0(980)K^*$ Decays }
\author{ Zhi-Qing
Zhang\footnote{Electronic address: zhangzhiqing@haut.edu.cn} , Jun-De Zhang}
\affiliation{ {\it \small  Department of Physics, Henan University
of Technology, Zhengzhou, Henan 450052, P.R.China} } %%

\begin{abstract}
\date{\today}

In this paper, the branching ratios and the direct CP-violating
asymmetries for decays $\bar B^0\to f_0(980)\bar K^{*0}$ and $B^-\to
f_0(980)K^{*-}$ by employing the perturbative QCD factorization
approach are studied. In the two-quark model supposition, $f_0(980)$ is commonly
viewed as a mixture of
 $s\bar{s}$ and $n\bar{n}(\equiv(u\bar{u}+d\bar{d})/\sqrt{2})$, that is
$|f_0(980)\ra = |s\bar s\ra\cos\theta+|n\bar n\ra\sin\theta$, where $\theta$ is the $f_0-\sigma$
mixing angle. We find that the non-factorizable $f_0$ emission type diagrams can give large contributions
to the final results, which are consistent with the present experimental data and the upper limit in the allowed
mixing angle ranges. We predict
that the direct CP asymmetry ${\cal A}^{dir}_{CP}(f_0(980)\bar K^{*0})$ is small, only a few percent, which
can be tested by future B factory experiments.
\end{abstract}

\pacs{13.25.Hw, 12.38.Bx, 14.40.Nd}
\vspace{1cm}

\maketitle

%=======================================================================
%                     Introduction
%=======================================================================

\section{Introduction}\label{intro}
In order to uncover the mysterious structure of the scalar meson $f_0(980)$, intensive studies have
been done since it was firstly observed in $\pi\pi$ scattering experiments \cite{Bellef0}.
There is still no consensus on the essential inner structure of $f_0(980)$. Some people consider
it as $q\bar{q}$ state \cite{nato}, or $qq\bar{qq}$ four-quark state \cite{jaffe}, other people think that
it is not made of one simple component but might have a more complex nature such as having
a $K\bar{K}$ component \cite{jwei,baru}, or mixing with glueball \cite{celenza,stro,close1}, or even superpositions
of the two- and four-quark states.

The B decays involved in the $f_0(980)$ in the final states are studied by employing various factorization approaches,
such as the generalization approach \cite{GMM}, the QCD factorization (QCDF) approach \cite{CYf0K,ccy,hycheng2},
the perturbative QCD (PQCD) approach \cite{Chenf0K1,Chenf0K2,wwang,zhangzq1}. In these calculation, the scalar meson
is usually viewed as a mixture of $s\bar{s}$ and $n\bar{n}(\equiv(u\bar{u}+d\bar{d})/\sqrt{2})$, that is
\be
|f_0(980)\ra = |s\bar s\ra\cos\theta+|n\bar n\ra\sin\theta,
\en
where $\theta$ is the $f_0-\sigma$ mixing angle. About the value of $\theta$, there are many discussions in
the phenomenal and experimental analyses \cite{theta,theta1}. But unfortunately it is difficult to find
a unique mixing angle to describe $f_0-\sigma$ mixing.

On the experimental side, for $f_0(980)$ emerging as a pole of the amplitude in the S wave \cite{kami}, many
channels such as $B\to f_0(980)K$ can be obtained by fitting  of Dalitz plots of the decays $B\to \pi^+\pi^-K$ and
$B\to \bar KKK$ and so on \cite{Bellef0,Bellef01,Bellef02, BaBarf0,BaBarf01,BaBarf02}. Although many such
decay channels that involved $f_0(980)$ in the final states have
been measured over the years, it has yet not been possible to account for the this scalar meson inner structure
, i.e. whether one deals with a two- or rather a four-quark composite, because there still lack precise and enough
data. For our considered decays, the measured values are\cite{pdg08}:
\be
Br(B^-\to f_0(980)K^{*-})&=&(10.4\pm2.6)\times 10^{-6},\\
Br(\bar B^0\to f_0(980)\bar K^{*0})&<&8.6\times 10^{-6}. \en
It is noticed that we have assumed $Br(f_0(980)\to\pi^+\pi^-)=0.50$ to obtain the upper experimental branching
ratios.

In this paper, we will study the branching ratios and the direct
CP asymmetries of $\bar B^0\to f_0(980)\bar K^{*0}$ and $B^-\to f_0(980)K^{*-}$
within perturbative QCD approach based on $k_T$ factorization.
In the following, $f_0(980)$ is denoted as $f_0$ in some places for convenience.
It is organized as follows. In Sect.\ref{proper}, the relevant decay constants
and light-cone distribution amplitudes of $B, f_0$ and $K^*$are  discussed.
In Sec.\ref{results}, we then analysis these decay channels using the pQCD approach.
The numerical results and the discussions are given
in section \ref{numer}. The conclusions are presented in the final part.

%=======================================================================
%       Physical properties of $f_0(980)$ and $f_0(1500)$
%=======================================================================

\section{decay constants and distribution amplitudes }\label{proper}

Now we present the wave functions to be used in the integration. For the wave function of the heavy B meson,
we take:
\be
\Phi_B(x,b)=
\frac{1}{\sqrt{2N_c}} (\psl_B +m_B) \gamma_5 \phi_B (x,b).
\label{bmeson}
\en
Here only the contribution of the Lorentz structure $\phi_B (x,b)$ is taken into account, since the contribution
of the second Lorentz structure $\bar \phi_B$ is numerically small \cite{cdlu} and has been neglected. For the
distribution amplitude $\phi_B(x,b)$ in Eq.(\ref{bmeson}), we adopt the model
\be
\phi_B(x,b)=N_Bx^2(1-x)^2\exp[-\frac{M^2_Bx^2}{2\omega^2_b}-\frac{1}{2}(\omega_bb)^2],
\en
where $\omega_b$ is a free parameter, and the value of the normalization factor is taken as $N_B=91.745$ for $\omega_b=0.4$
in numerical calculations.

In  the two-quark model, the vector decay constant $f_{f_0}$ and the scalar  decay constant
$\bar {f}_{f_0}$ for scalar meson $f_0$
can  be defined as:
\be
\langle f_0(p)|\bar q_2\gamma_\mu q_1|0\ra&=&f_{f_0}p_\mu,
\en
\be
\langle f_0(p)|\bar q_2q_1|0\ra=m_{f_0}\bar {f}_{f_0}. \label{fbar}
\en

Owing to charge conjugation invariance or the G parity conservation, the neutral scalar meson $f_0$
cannot be produced via the vector current, so $f_{f_0}=0$. Taking the mixing into account,
Eq.(\ref{fbar}) is changed to
\be \langle f_0^n|\bar
dd|0\ra=\langle f_0^n|\bar uu|0\ra=\frac{1}{\sqrt 2}m_{f_0}\tilde
f^n_{f_0},\,\,\,\, \langle f_0^n|\bar ss|0\ra=m_{f_0}\tilde
f^s_{f_0}.
\en
Using the QCD sum-rule method, one can find that the
scale-dependent scalar decay constants $\tilde f_{f_0}^n$ and $\tilde f_{f_0}^s$
are very close\cite{ccy}. So $\tilde
f_{f_0}^n=\tilde f_{f_0}^s$ is assumed and we denote them as $\bar f_{f_0}$ in the
following.

The light-cone distribution amplitudes (LCDAs) for the  scalar meson
$f_0$ can be written as: \be \langle f_0(p)|\bar q_1(z)_l
q_2(0)_j|0\rangle &=&\frac{1}{\sqrt{2N_c}}\int^1_0dx \; e^{ixp\cdot
z}\non && \times \{ p\sl\Phi_{f_0}(x)
+m_{f_0}\Phi^S_{f_0}(x)+m_{f_0}(n\sl_+n\sl_--1)\Phi^{T}_{f_0}(x)\}_{jl},\quad\quad\label{LCDA}
\en here $n_+$ and $n_-$ are light-like vectors:
$n_+=(1,0,0_T),n_-=(0,1,0_T)$, and $n_+$ is parallel with the moving
direction of the scalar meson $f_0$. The normalization can be
related to the decay constants: \be \int^1_0
dx\Phi_{f_0}(x)=\int^1_0
dx\Phi^{T}_{f_0}(x)=0,\,\,\,\,\,\,\,\int^1_0
dx\Phi^{S}_{f_0}(x)=\frac{\bar f_{f_0}}{2\sqrt{2N_c}}\;. \en The
twist-2 LCDA can be expanded in the Gegenbauer polynomials: \be
\Phi_{f_0}(x,\mu)&=&\frac{1}{2\sqrt{2N_c}}\bar
f_{f_0}(\mu)6x(1-x)\sum_{m=1}^\infty B_m(\mu)C^{3/2}_m(2x-1), \en
the values for Gegenbauer moments $B_1, B_3$ have been calculated in
 \cite{ccy} as: \be B_1=-0.78\pm0.08,\quad\quad B_3=0.02\pm0.07. \en
These values are taken at $\mu=1$ GeV and the even Gegenbauer moments
vanish.

As for the twist-3 distribution amplitudes $\Phi_{f_0}^S$ and $\Phi_{f_0}^T$,
they have not been studied in the literature, so we adopt the asymptotic form :
\be
\Phi^S_{f_0}&=& \frac{1}{2\sqrt {2N_c}}\bar f_{f_0},\,\,\,\,\,\,\,\Phi_{f_0}^T=
\frac{1}{2\sqrt {2N_c}}\bar f_{f_0}(1-2x).
\en

For our considered
decays, the vector meson $K^*$ is longitudinally polarized. The
longitudinal polarized component of the wave function is given as:
\be
\Phi_{K^*}=\frac{1}{\sqrt{2N_c}}\left\{\esl\left[m_{K^*}\Phi_{K^*}(x)+\pvsl\Phi_{K^*}^t(x)\right]+m_{K^*}\Phi^s_{K^*}(x)\right\},
\en
where the first term is the leading twist wave function (twist-2),
while the second and third term are sub-leading twist (twist-3) wave
functions. They can be parameterized as:
\be
 \Phi_{K^*}(x) &=&  \frac{f_{K^*}}{2\sqrt{2N_c} }
    6x (1-x)
    \left[1+a_{1K^*}C^{3/2}_1(2x-1)+a_{2K^*}C^{3/2}_2(2x-1)\right],\label{piw1}
\en
\be
 \Phi^t_{K^*}(x) =   \frac{3f^T_{K^*}}{2\sqrt{2N_c}}(1-2x),\quad \Phi^s_{K^*}(x)
 =\frac{3f^T_{K^*}}{2\sqrt{2N_c}}(2x-1)^2,\label{piw}
\en
where the Gegenbauer moments $a_{1K^*}=0.03, a_{2K^*}=0.11$ \cite{pball} and the Gegenbauer polynomials $C^{\nu}_n(t)$ are given as:
\be
C^{3/2}_1(t)&=&3t, \qquad C^{3/2}_2(t)=\frac{3}{2}(5t^2-1),\\
C^{3/2}_3(t)&=&\frac{5}{2}t(7t^2-3).\label{eq:c124}
\en
%===========================================================================
%                    Decay amplitudes in PQCD approach
%============================================================================

\section{ the perturbative QCD  calculation} \label{results}

Under the two-quark model for the scalar meson $f_0$ supposition, we
would like to use pQCD approach to study B decays into $f_{0}$
and $K^*$.
The decay amplitude can be conceptually written as the convolution,
\be
{\cal A}(B \to  f_0K^*)\sim \int\!\! d^4k_1
d^4k_2 d^4k_3\ \mathrm{Tr} \left [ C(t) \Phi_B(k_1) \Phi_{f_0}(k_2)
\Phi_{K^*}(k_3) H(k_1,k_2,k_3, t) \right ], \label{eq:con1}
\en
where $k_i$'s are momenta of anti-quarks included in each mesons, and
$\mathrm{Tr}$ denotes the trace over Dirac and color indices. $C(t)$
is the Wilson coefficient, which results from the radiative
corrections at short distance. In the above convolution, $C(t)$
includes the harder dynamics at larger scale than $M_B$ scale and
describes the evolution of local four-Fermi operators from $m_W$ (the
$W$ boson mass) down to $t\sim\mathcal{O}(\sqrt{\bar{\Lambda} M_B})$
scale, where $\bar{\Lambda}\equiv M_B -m_b$. The function
$H(k_1,k_2,k_3,t)$ describes the four-quark operator and the
spectator quark connected by
 a hard gluon whose $q^2$ is in the order
of $\bar{\Lambda} M_B$, and includes the
$\mathcal{O}(\sqrt{\bar{\Lambda} M_B})$ hard dynamics. Therefore,
this hard part $H$ can be perturbatively calculated.

Since the b quark is rather heavy we consider the $B$ meson at rest
for simplicity. It is convenient to use light-cone coordinates $(p^+,
p^-, {\bf p}_T)$ to describe the meson's momenta by
\be p^\pm =
\frac{1}{\sqrt{2}} (p^0 \pm p^3), \quad {\rm and} \quad {\bf p}_T =
(p^1, p^2). \en Using these coordinates the $B$ meson and the two
final state meson momenta can be written as
\be
P_B =\frac{m_B}{\sqrt{2}} (1,1,{\bf 0}_T), \quad
P_{2} =\frac{m_B}{\sqrt{2}}(1-r^2_{K^*},r^2_{f_0},{\bf 0}_T), \quad
P_{3} =\frac{m_B}{\sqrt{2}} (r^2_{K^*},1-r^2_{f_0},{\bf 0}_T),
\en
respectively. Here we have the mass ratios
\be
r_{K^*}=m_{K^*}/m_B,\quad\quad\quad r_{f_0}=m_{f_0}/m_B.\label{rmass}
\en
Putting the anti-quark momenta in $B$,
$f_0$, $K^*$ mesons as $k_1$, $k_2$, and $k_3$, respectively, we can
choose
\be
k_1 = (x_1 P_1^+,0,{\bf k}_{1T}), \quad
k_2 = (x_2 P_2^+,0,{\bf k}_{2T}), \quad
k_3 = (0, x_3 P_3^-,{\bf k}_{3T}).
\en
For these considered decay channels, the integration over $k_1^-$,
$k_2^-$, and $k_3^+$ in eq.(\ref{eq:con1}) will lead to
\be
 {\cal
A}(B \to f_0 K^*) &\sim &\int\!\! d x_1 d x_2 d x_3 b_1 d b_1 b_2 d
b_2 b_3 d b_3 \non && \cdot \mathrm{Tr} \left [ C(t) \Phi_B(x_1,b_1)
\Phi_{f_0}(x_2,b_2) \Phi_{K^*}(x_3, b_3) H(x_i, b_i, t) S_t(x_i)\,
e^{-S(t)} \right ], \quad \label{eq:a2}
\en
where $b_i$ is the
conjugate space coordinate of $k_{iT}$, and $t$ is the largest
energy scale in function $H(x_i,b_i,t)$.
In order to smear the end-point singularity on $x_i$,
the jet function $S_t(x)$ \cite{li02}, which comes from the
resummation of the double logarithms $\ln^2x_i$, is used.
The last term $e^{-S(t)}$ in Eq.(\ref{eq:a2}) is the Sudakov form factor, which suppresses
the soft dynamics effectively \cite{soft}.

 For the considered decays, the related weak effective
Hamiltonian $H_{eff}$ can be written as \cite{buras96}
\be
\label{eq:heff} {\cal H}_{eff} = \frac{G_{F}} {\sqrt{2}} \,
\sum_{q=u,c}V_{qb} V_{qs}^*\left[ \left (C_1(\mu) O_1^q(\mu) +
C_2(\mu) O_2^q(\mu) \right) \sum_{i=3}^{10} C_{i}(\mu) \,O_i(\mu)
\right] \; ,
\en
with the Fermi constant $G_{F_0}=1.166 39\times
10^{-5} GeV^{-2}$, and the CKM matrix elements V. We specify below
the operators in ${\cal H}_{eff}$ for $b \to s$ transition:
\be
\begin{array}{llllll}
O_1^{u} & = &  \bar s_\alpha\gamma^\mu L u_\beta\cdot \bar
u_\beta\gamma_\mu L b_\alpha\ , &O_2^{u} & = &\bar
s_\alpha\gamma^\mu L u_\alpha\cdot \bar
u_\beta\gamma_\mu L b_\beta\ , \\
O_3 & = & \bar s_\alpha\gamma^\mu L b_\alpha\cdot \sum_{q'}\bar
 q_\beta'\gamma_\mu L q_\beta'\ ,   &
O_4 & = & \bar s_\alpha\gamma^\mu L b_\beta\cdot \sum_{q'}\bar
q_\beta'\gamma_\mu L q_\alpha'\ , \\
O_5 & = & \bar s_\alpha\gamma^\mu L b_\alpha\cdot \sum_{q'}\bar
q_\beta'\gamma_\mu R q_\beta'\ ,   & O_6 & = & \bar
s_\alpha\gamma^\mu L b_\beta\cdot \sum_{q'}\bar
q_\beta'\gamma_\mu R q_\alpha'\ , \\
O_7 & = & \frac{3}{2}\bar s_\alpha\gamma^\mu L b_\alpha\cdot
\sum_{q'}e_{q'}\bar q_\beta'\gamma_\mu R q_\beta'\ ,   & O_8 & = &
\frac{3}{2}\bar s_\alpha\gamma^\mu L b_\beta\cdot
\sum_{q'}e_{q'}\bar q_\beta'\gamma_\mu R q_\alpha'\ , \\
O_9 & = & \frac{3}{2}\bar s_\alpha\gamma^\mu L b_\alpha\cdot
\sum_{q'}e_{q'}\bar q_\beta'\gamma_\mu L q_\beta'\ ,   & O_{10} & =
& \frac{3}{2}\bar s_\alpha\gamma^\mu L b_\beta\cdot
\sum_{q'}e_{q'}\bar q_\beta'\gamma_\mu L q_\alpha'\ ,
\label{eq:operators} \end{array}
\en
where $\alpha$ and $\beta$ are
the $SU(3)$ color indices; $L$ and $R$ are the left- and
right-handed projection operators with $L=(1 - \gamma_5)$, $R= (1 +
\gamma_5)$. The sum over $q'$ runs over the quark fields that are
active at the scale $\mu=O(m_b)$, i.e., $(q'\epsilon\{u,d,s,c,b\})$.

%===========================================================================
%                  Numerical results and discussions
%============================================================================

\begin{figure}[t,b]
\vspace{-3cm} \centerline{\epsfxsize=16 cm \epsffile{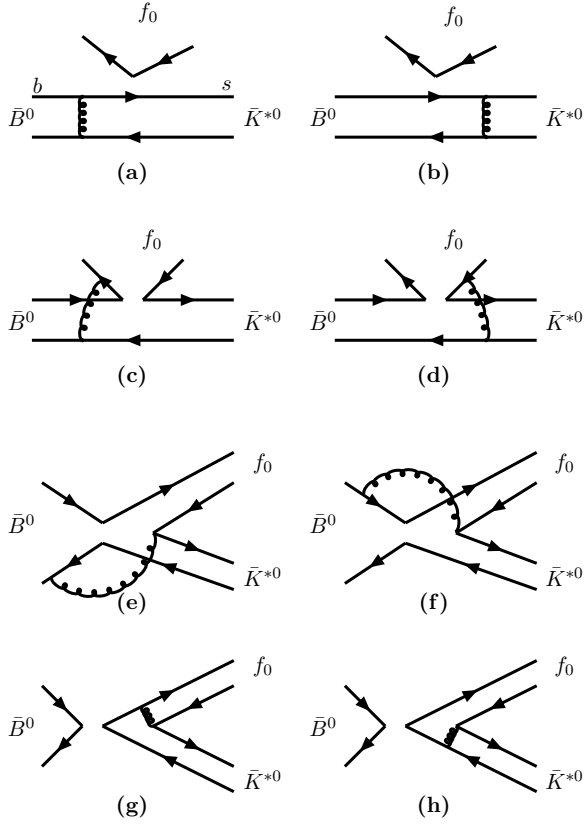}}
\vspace{-9cm} \caption{ Diagrams contributing to the decay $\bar{B}^0\to f_0(980)
\bar K^{*0}$ .}
 \label{Fig1}
\end{figure}

In the following, we take the $\bar{B}^0\to f_0 \bar{K}^{*0}$ decay channel as an example to expound.
There are 8 type diagrams contributing to this decay, as illustrated in Fig.\ref{Fig1}.
For the factorizable emission diagrams (a) and (b), operators $O_{1-4,9,10}$ are
$(V-A)(V-A)$ currents, and the operators $O_{5-8}$ have the
structure of $(V-A)(V+A)$, the sum of the their amplitudes are
written as $F_{eK^*}$ and $F_{eK^*}^{P1}$, respectively.
For $ \langle f_0 |\bar{q}\gamma_\mu q| 0\rangle =0 $, one then finds that
\be
F_{eK^*}&=&F_{eK^*}^{P1}=0 . \label{eq:as}
\en

In order to get the right flavor and color structure for factorization to work,  a Fierz transformation
for the $(V-A)(V+A)$ operators may sometimes be needed and then the corresponding amplitude is

\begin{eqnarray} F^{P2}_{eK^*} &=&-16 \pi C_F m_B^4 r_{f_0}
{\bar f_{f_0}}\int_0^1 dx_1 dx_3 \int_0^{\infty} b_1db_1\, b_3db_3\,
\Phi_B(x_1,b_1)
\nonumber \\
& &\times \bigg\{ \left[ \Phi_{K^*}(x_3)-r_{K^*} x_3 \left(
\Phi_{K^*}^s(x_3) + \Phi_{K^*}^t(x_3) \right) -2 r_{K^*}
\Phi_{K^*}^s(x_3)
 \right]\nonumber\\
&&\;\;\;\;  \times E_{ei}(t) h_{e}(x_1,x_3,b_1,b_3) - 2 r_{K^*}
 \Phi_{K^*}^s(x_3)
E_{ei}(t') h_{e}(x_3,x_1,b_3,b_1) \bigg\}\;,\label{first1}
\end{eqnarray}
where $C_F=4/3$ is the group factor of the $SU(3)_c$ gauge group. The functions $E_{ei}(t^{(\prime)})$ and $h_e$
including the Sudakov factor and jet function have the same definition as those in Ref. \cite{wwang,zhangzq1}.

For the non-factorizable diagrams 1(c) and 1(d), all three meson
wave functions are involved. The integration of $b_3$ can be
performed using the $\delta$ function $\delta(b_3-b_2)$, leaving only
integration of $b_1$ and $b_2$. For the $(V-A)(V-A)$, $(V-A)(V+A)$,
$(S-P)(S+P)$ operators, the results are
\begin{eqnarray} {\cal M}_{eK^*} &=& 32 \pi C_Fm_B^4
/\sqrt{2N_C}\int_0^1 dx_1dx_2dx_3 \int_0^{\infty} b_1 db_1\, b_2
db_2\,\Phi_B(x_1,b_1) \Phi_{f_0}(x_2) \non
&&\bigg\{[(x_2-1)\Phi_{K^*}(x_3)-r_{K^*}x_3(\Phi_{K^*}^s(x_3)+\Phi_{K^*}^t(x_3))]
E'_{ei}(t) h_n(x_1,1-x_2,x_3,b_1,b_2)
\nonumber \\
&&+[(x_2+x_3)\Phi_{K^*}(x_3) +r_{K^*} x_3 \left(
\Phi_{K^*}^s(x_3)-\Phi_{K^*}^t(x_3)
       \right]
E'_{ei}(t') h_n(x_1,x_2,x_3,b_1,b_2) \bigg\},\;\;\;\quad \label{first2}
\end{eqnarray}
%%%%
\begin{eqnarray}
{\cal M}^{P1}_{eK^*}&=& 32 \pi C_Fm_B^4r_{f_0} /\sqrt{2N_C}\int_0^1
dx_1dx_2dx_3 \int_0^{\infty} b_1 db_1\, b_2 db_2\,\Phi_B(x_1,b_1)
\nonumber \\
& &\times \bigg\{ E'_{ei}(t) h_n(x_1,1-x_2,x_3,b_1,b_2)\left[
(x_2-1)\Phi_{K^*}(x_3) \left( \Phi_{f_0}^S(x_2) - \Phi_{f_0}^T(x_2)
\right)
\right.\nonumber \\
& &\;\;\;\;\;\;-r_{K^*}(x_2-1)\left(
\Phi_{K^*}^s(x_3)+\Phi_{K^*}^t(x_3) \right) \left( \Phi_{f_0}^S(x_2)-
\Phi_{f_0}^T(x_2)
\right)\nonumber\\
& &\;\;\;\;\;\;\left. + r_{K^*} x_3 \left(
\Phi_{K^*}^s(x_3)-\Phi_{K^*}^t(x_3) \right) \left( \Phi_{f_0}^S(x_2)+
\Phi_{f_0}^T(x_2) \right)\right]
\nonumber \\
& & \;\;\;\;\;\;+  E'_{ei}(t') h_n(x_1,x_2,x_3,b_1,b_2) \left[
x_2\Phi_{K^*}(x_3) \left( \Phi_{f_0}^S(x_2) + \Phi_{f_0}^T(x_2)
\right) \right.\nonumber \\
&&\;\;\;\;\; -r_{K^*}x_2\left( \Phi_{K^*}^s(x_3)+\Phi_{K^*}^t(x_3)
\right) \left( \Phi_{f_0}^S(x_2)+ \Phi_{f_0}^T(x_2)
\right)\nonumber\\
& &\;\;\;\;\;\;\left. - r_{K^*} x_3 \left(
\Phi_{K^*}^s(x_3)-\Phi_{K^*}^t(x_3) \right) \left( \Phi_{f_0}^S(x_2)-
\Phi_{f_0}^T(x_2) \right)\right]\bigg\}\;,
\end{eqnarray}
\begin{eqnarray}
{\cal M}^{P2}_{eK^*} &=& -32 \pi C_Fm_B^4 /\sqrt{2N_C}\int_0^1
dx_1dx_2dx_3 \int_0^{\infty} b_1 db_1\, b_2
db_2\,\Phi_B(x_1,b_1)\Phi_{f_0}(x_2)
\nonumber \\
& &\times \bigg\{ \left[ (-x_2+x_3+1)\Phi_{K^*}(x_3)
+r_{K^*}x_3\left(\Phi_{K^*}^s(x_3)-\Phi_{K^*}^t(x_3) \right)\right]
\nonumber \\
& &\times E'_{ei}(t) h_n(x_1,1-x_2,x_3,b_1,b_2)-E'_{ei}(t')h_n(x_1,x_2,x_3,b_1,b_2)
\nonumber \\
& &
\times\left[x_2\Phi_{K^*}(x_3)+r_{K^*}
x_3(\Phi_{K^*}^s(x_3)+\Phi_{K^*}^t(x_3))\right] \bigg\}\;.
\end{eqnarray}

For the non-factorizable annihilation diagrams (e) and (f), again
all three wave functions are involved. $M_{aK^*}$, $M_{aK^*}^{P1}$ and $M_{aK^*}^{P2}$
describe the contributions from the $(V-A)(V-A)$, $(V-A)(V+A)$ and
$(S-P)(S+P)$ type operators, respectively,
\be
{\cal M}_{aK^*} &=&
32\pi C_Fm_B^4 /\sqrt{2N_C}\int_0^1 dx_1dx_2dx_3 \int_0^{\infty} b_1
db_1\, b_2 db_2\,\Phi_B(x_1,b_1)\left\{ \left[ x_2\Phi_{K^*}(x_3) \Phi_{f_0}(x_2)
\right.\right.\nonumber \\
& &\left.\left.
+r_{K^*}r_{f_0} \Phi_{f_0}^T(x_2) \left(
(x_2+x_3-1)\Phi_{K^*}^s(x_3)+(x_2-x_3-1) \Phi_{K^*}^t(x_3)
\right)\right.\right.\nonumber\\
& &\left.\left. - r_{K^*}r_{f_0} \Phi_{f_0}^S(x_2) \left(
(x_2-x_3+3)\Phi_{K^*}^s(x_3)+(x_2+x_3-1) \Phi_{K^*}^t(x_3)
\right)\nonumber\right] E'_{ai}(t)
\right.\nonumber \\
& &\left.\times h_{na}(x_1,x_2,x_3,b_1,b_2)+  E'_{ai}(t') h'_{na}(x_1,x_2,x_3,b_1,b_2)\left[
(x_3-1)\Phi_{K^*}(x_3) \Phi_{f_0}(x_2)
\right.\right.\nonumber \\
& &\left.\left.+r_{K^*}r_{f_0} \Phi_{f_0}^S(x_2) \left(
(x_2-x_3+1)\Phi_{K^*}^s(x_3)-(x_2+x_3-1) \Phi_{K^*}^t(x_3)
\right)\right.\right.\nonumber\\
& &\left.\left. - r_{K^*}r_{f_0} \Phi_{f_0}^T(x_2) \left(
(1-x_2-x_3)\Phi_{K^*}^s(x_3)+(1+x_2-x_3) \Phi_{K^*}^t(x_2)
\right)\right] \right\}\;,
\en
\be
 {\cal M}^{P1}_{aK^*} &=& 32 \pi C_Fm_B^4 /\sqrt{2N_C}\int_0^1
dx_1dx_2dx_3 \int_0^{\infty} b_1 db_1\, b_2 db_2\,\Phi_B(x_1,b_1)
\nonumber \\
&&\times \bigg\{ \left[r_{K^*} (1+x_3)\Phi_{f_0}(x_2)
(\Phi_{K^*}^t(x_3)+\Phi_{K^*}^s(x_3))
-r_{f_0}(x_2-2)\Phi_{K^*}(x_3)\left(\Phi_{f_0}^S(x_2)\right.\right.\nonumber \\
&&\;\;\left.\left.-\Phi_{f_0}^T(x_2)\right)\right]E'_{ai}(t) h_{na}(x_1,x_2,x_3,b_1,b_2)-  E'_{ai}(t') h'_{na}(x_1,x_2,x_3,b_1,b_2)\left[r_{K^*}
\right.\nonumber \\
&& \left.\times  (x_3-1)\Phi_{f_0}(x_2)
(\Phi_{K^*}^t(x_3)+\Phi_{K^*}^s(x_3))-r_{f_0}x_2\Phi_{K^*}(x_3)(\Phi_{f_0}^S(x_2)-\Phi_{f_0}^T(x_2))
\right]
\bigg\},\;\;\;\;\;\;
\en
\begin{eqnarray}
 {\cal M}^{P2}_{aK^*} &=& 32 \pi C_Fm_B^4 /\sqrt{2N_C}\int_0^1
dx_1dx_2dx_3 \int_0^{\infty} b_1 db_1\, b_2 db_2\,\Phi_B(x_1,b_1)
\nonumber \\
&&\times \bigg\{ \left[(x_3-1)\Phi_{f_0}(x_2)
\right.\Phi_{K^*}(x_3)+4r_{K^*}r_{f_0}\Phi^S_{f_0}(x_2)\Phi_{K^*}^s(x_3)
+r_{K^*}r_{f_0}\left((x_2-x_3-1)\right.\nonumber
\\&&\;\times\left(\Phi_{K^*}^s(x_3)\Phi_{f_0}^S(x_2)-\Phi_{K^*}^t(x_3)\Phi_{f_0}^T(x_2)\right.)\left.
+(x_2+x_3-1)\left(\Phi_{K^*}^s(x_3)\Phi_{f_0}^T(x_2)\right.\right.\nonumber \\
&&\left.\left.-\Phi_{K^*}^t(x_3)\Phi_{f_0}^S(x_2)\right)\right] E'_{ai}(t) h_{na}(x_1,x_2,x_3,b_1,b_2)+E'_{ai}(t') h'_{na}(x_1,x_2,x_3,b_1,b_2)
\nonumber \\
&& \;\;\times \left[x_2 \Phi_{f_0}(x_2)
\Phi_{K^*}(x_3)-x_2r_{K^*}r_{f_0}(\Phi_{f_0}^S(x_2)-\Phi_{f_0}^T(x_2))(\Phi_{K^*}^s(x_3)+\Phi_{K^*}^t(x_3))
\right.\nonumber \\ &&\;\;\;\left. -r_{K^*}r_{f_0}
(1-x_3)(\Phi_{f_0}^S(x_2)+\Phi_{f_0}^T(x_2))(\Phi_{K^*}^s(x_3)-\Phi_{K^*}^t(x_3))
\right]\bigg\}\;.
\end{eqnarray}

The factorizable annihilation diagrams (g) and (h) involve only two
final state mesons' wave functions. There are also three kinds of decay
amplitudes for these two diagrams. $F_{aK^*}$ is for $(V-A)(V-A)$
type operators, $F_{aK^*}^{P1}$ is for $(V-A)(V+A)$ type operators,
while $F_{aK^*}^{P2}$ is for $(S-P)(S+P)$ type operators:
\begin{eqnarray}
F_{aK^*} &=&F^{P1}_{aK^*}= -8 \pi C_F m_B^4f_B \int_0^1 dx_2 dx_3
\int_0^{\infty} b_2db_2\, b_3db_3\left\{\left[(x_3-1)\Phi_{K^*}(x_3)\Phi_{f_0}(x_2)
\right.\right.\nonumber \\
&& \left.\left. - 2r_{K^*}
r_{f_0}(x_3-2) \Phi_{K^*}^s(x_3)\Phi_{f_0}^S(x_2) -2r_{K^*} r_{f_0}x_3
\Phi_{K^*}^t(x_3)\Phi_{f_0}^S(x_2) \right]\right.
\nonumber \\
&&\left.\times E_{ai}(t) h_{a}(x_2,1-x_3, b_2, b_3)+E_{ai}(t') h_{a}(1-x_3,x_2, b_3, b_2)
\right.\nonumber\\
&&
\left.\times [x_2\Phi_{K^*}(x_3)\Phi_{f_0}(x_2)-2r_{K^*}
r_{f_0}\Phi_{K^*}^s(x_3)((x_2+1)\Phi_{f_0}^S(x_2)-(x_2-1)\Phi_{f_0}^T)]\right\}, \en

\be
F^{P2}_{aK^*} &=& 16 \pi C_F m_B^4f_B \int_0^1 dx_2dx_3 \int_0^{\infty} b_2db_2\, b_3db_3
\nonumber\\ &&
\times\bigg\{[r_{K^*}(x_3-1)\Phi_{f_0}(x_2)(\Phi_{K^*}^s(x_3)-\Phi_{K^*}^t(x_3))
+2r_{f_0}\Phi_{K^*}(x_3)\Phi_{f_0}^S(x_2)]\nonumber\\ &&\times E_{ai}(t) h_{a}(x_2,1-x_3,
b_2, b_3) -E_{ai}(t') h_{a}(1-x_3,x_2, b_3,b_2)\nonumber\\ &&
\times [2r_{K^*}\Phi_{K^*}^s(x_3)
\Phi_{f_0}(x_2)-r_{f_0}x_2\Phi_{K^*}(x_3)
(\Phi_{f_0}^T(x_2)+\Phi_{f_0}^S(x_2))]\bigg\}.
\label{last1}
\en

If we exchange the $K^*$ and $f_0$ in Fig.\ref{Fig1}, the corresponding expressions of amplitudes for new
diagrams will be similar with those as given in Eqs.(\ref{first2}-\ref{last1}) and can be obtained by the
replacements:
\be
\Phi_{f_0}(x)\longleftrightarrow \Phi_{K^*}(x), \Phi^{S}_{f_0}(x)\longleftrightarrow \Phi^{s}_{K^*}(x),
\Phi^{T}_{f_0}(x)\longleftrightarrow \Phi^{t}_{K^*}(x), r_{f_0}\longleftrightarrow r_{K^*},
\en
since the wave functions for the mesons $f_0(980)$ and $K^*$ have exactly the same form.
The only difference is some normalization
constants for the different twist distribution amplitudes. That is, the factorization formulae for
(a) and (b) in the new diagrams amplitudes are written as:
\begin{eqnarray}
F_{ef_0} &=& -8 \pi C_F m_B^4 f_{K^*}\int_0^1 dx_1 dx_2
\int_0^{\infty} b_1db_1\, b_2db_2\, \Phi_B(x_1,b_1)
\nonumber \\
& & \times \bigg\{ \left[ (1+x_2)\Phi_{f_0}(x_2)-r_{f_0}(1-2x_2)
\left( \Phi_{f_0}^S(x_2)-\Phi_{f_0}^T(x_2) \right) \right] E_{ei}(t)
h_{e}(x_1,x_2,b_1,b_2)
\nonumber\\
& &\;\;\;\;\;\; -2r_{f_0} \Phi_{f_0}^S({x_2})E_{ei}(t')
h_{e}(x_2,x_1,b_2,b_1) \bigg\} \;,
\label{f0from}
\end{eqnarray}
%%%%
%%%%
\begin{eqnarray}
 F^{P2}_{ef_0}&=&0.
 \end{eqnarray}

Since we have chosen the momentum fraction at the anti-quark, we should use $\Phi^{(S,T)}_{f_0}(1-x)$ and
$\Phi^{(s,t)}_{K^*}(1-x)$ for the mesons $f_0(980)$ and $K^*$ in the calculation.
But for simplicity, we use $\Phi^{(S,T)}_{f_0}(x)$ and $\Phi^{(s,t)}_{K^*}(x)$ to denote
$\Phi^{(S,T)}_{f_0}(1-x)$ and $\Phi^{(s,t)}_{K^*}(1-x)$ in the upper formulae.

Combining the contributions from different diagrams, the total decay
amplitudes for the decays $\bar{B}^0\to f_0 \bar{K}^{*0}$ and $B^-\to f_0 K^{*-}$ can be written as:
\be
{\cal M}( f_0K^{0} ) &=&\xi_uM_{eK^*}C_2F_1(\theta)-\xi_t\left\{F_{ef_0}(a_4-\frac{a_{10}}{2})F_1(\theta)+[F^{P2}_{ef_0}F_1(\theta)+
F^{P2}_{eK^*}F_2(\theta)](a_6-\frac{a_8}{2})\right.\nonumber\\ &&\left.+[M_{ef_0}(C_3-\frac{C_9}{2})+M_{eK^*}(2C_4+\frac{C_{10}}{2})]F_1(\theta)
+M_{eK^*}(C_3+C_4-\frac{C_9}{2}-\frac{C_{10}}{2})\right.\nonumber\\ &&\left.\times F_2(\theta)+
[M^{P1}_{ef_0}F_1(\theta)+M^{P1}_{eK^*}F_2(\theta)](C_5-\frac{C_7}{2})+M^{P2}_{eK^*}\left[(2C_6+\frac{C_8}{2})F_1(\theta)
\right.\right.\nonumber\\ &&\left.\left.+(C_6-\frac{C_8}{2})F_2(\theta)\right]+\left[M_{af_0}F_1(\theta)
+M_{aK^*}F_2(\theta)\right](C_3-\frac{C_9}{2})+\left[M^{P1}_{af_0}F_1(\theta)
\right.\right.\nonumber\\ &&\left.\left.+M^{P1}_{aK^*}F_2(\theta)\right](C_5-\frac{C_7}{2})
+[F_{af_0}F_1(\theta)+F_{aK^*}F_2(\theta)](a_4-\frac{a_{10}}{2})\right.\nonumber\\ &&\left.
+[F^{P2}_{af_0}F_1(\theta)+F^{P2}_{aK^*}F_2(\theta)](a_6-\frac{a_{8}}{2})\right\}
,\label{eq:kst0}
\en
\be
{\cal M}( f_0K^{*-} ) &=&\xi_u[(F_{ef_0}a_1+M_{eK^*}C_2+M_{ef_0}C_1+M_{af_0}C_1+F_{af_0}a_1)F_1(\theta)+\left(M_{aK^*}\right.\nonumber\\ &&\left.\times C_1+F_{aK^*}a_1\right)F_2(\theta)]
-\xi_t\left\{F_{ef_0}(a_4+a_{10})F_1(\theta)+F^{P2}_{ef_0}F_1(\theta)(a_6+a_8)
\right.\nonumber\\ &&\left.+F^{P2}_{eK^*}F_2(\theta)(a_6-\frac{a_8}{2})+[M_{ef_0}(C_3+C_9)+M_{eK^*}(2C_4+\frac{C_{10}}{2})]F_1(\theta)
\right.\nonumber\\ &&\left.+M_{eK^*}(C_3+C_4-\frac{C_9}{2}-\frac{C_{10}}{2})F_2(\theta)+
M^{P1}_{ef_0}F_1(\theta)(C_5+C_7)+M^{P1}_{eK^*}
\right.\nonumber\\&&\left. \times F_2(\theta)(C_5-\frac{C_7}{2})+M^{P2}_{eK^*}[(2C_6+\frac{C_8}{2})F_1(\theta)+(C_6-\frac{C_8}{2})
F_2(\theta)]\right.\nonumber\\&&\left.+[M_{af_0}F_1(\theta)+M_{aK^*}F_2(\theta)](C_3+C_9)+[M^{P1}_{af_0}F_1(\theta)+M^{P1}_{aK^*}F_2(\theta)]\right.\nonumber\\ &&\left.
\times(C_5+C_7)+[F_{af_0}F_1(\theta)+F_{aK^*}F_2(\theta)](a_4+a_{10})\right.\nonumber\\&&\left.+[F^{P2}_{af_0}F_1(\theta)+F^{P2}_{aK^*}F_2(\theta)](a_6+a_{8})\right\}
,\label{eq:kstf}
\en
where $\xi_u=V_{ub}V^*_{us}, \xi_t=V_{tb}V^*_{ts}$ and $F_1(\theta)=\sin\theta/\sqrt2, F_2(\theta)=\cos\theta$. The combinations of the Wilson coefficients are defined as usual
\cite{AKL}:
 \be
a_{1}(\mu)&=&C_2(\mu)+\frac{C_1(\mu)}{3}, \quad
a_2(\mu)=C_1(\mu)+\frac{C_2(\mu)}{3},\non
a_i(\mu)&=&C_i(\mu)+\frac{C_{i+1}(\mu)}{3},\quad
i=3,5,7,9,\non
a_i(\mu)&=&C_i(\mu)+\frac{C_{i-1}(\mu)}{3},\quad
i=4, 6, 8, 10.\label{eq:aai} \en

\section{Numerical results and discussions} \label{numer}

\begin{table}
\caption{Input parameters used in the numerical calculation\cite{ccy,pdg08}.}\label{para}
\begin{center}
\begin{tabular}{c |ccc}
\hline \hline
 Masses &$m_{f_0}=0.980 \mbox{ GeV}$,   &$ m_{K^*}=0.892 \mbox{ GeV}$,\\
  & $ M_B = 5.28 \mbox{ GeV}$,&\\
 \hline
  Decay constants &$f_B = 0.19 \mbox{ GeV}$,  & $f_{f_0} = 0.37 \mbox{ GeV}$,\\ &$f_{K^*} = 0.217
 \mbox{ GeV}$,&$f^T_{K^*}=0.185 \mbox{ GeV}$,\\
 \hline
Lifetimes &$\tau_{B^\pm}=1.638\times 10^{-12}\mbox{ s}$, &
$\tau_{B^0}=1.530\times 10^{-12}\mbox{ s}$,\\
 \hline
$CKM$ &$V_{tb}=1.0$, & $V_{ts}=-0.0387$,\\
 &$V_{us}=0.2255$,& $V_{ub}=0.00393e^{-i60^{\circ}}$. \\
\hline \hline
\end{tabular}
\end{center}
\end{table}

In the numerical calculation, we will use the input parameters as listed
in Table~\ref{para}.

From Eq.(\ref{f0from}), we can find the numerical values of the corresponding form factor
$F^{\bar B^0\to f_0(d\bar d)}_0$at maximal recoiling:
\be
F^{\bar B^0\to f_0(d\bar d)}_0=0.31,
\en
which is smaller than $F^{\bar B^0\to f_0(980)(d\bar d)}=0.47$ \cite{wwang}, for using different values
for the threshold parameters $c$ in the jet function.

In the B-rest frame, the decay rate of $B\to f_0(980)K^*$ can be written as:
\be
\Gamma=\frac{G_F^2}{32\pi m_B}|{\cal M}|^2(1-r^2_{f_0}-r^2_{K^*}),
\en
where $r_{f_0}, r_{K^*}$ have been defined in Eq.(\ref{rmass}) and ${\cal M}$ is the total decay amplitude of
$B\to f_0(980)K^*$, which has been given  in  section \ref{results}.

Using the wave functions as specified in the previous section and the input parameters
listed in Table~\ref{para}, it is straightforward to calculate the
CP-averaged branching ratios for the considered decays.

If $f_0(980)$ is purely composed of $s\bar s$, the branching ratios of $B\to f_0(980)K^*$ are:
\be
Br(\bar B^0\to f_0(980)\bar K^{*0})=(14.0^{+1.5+1.2+2.9}_{-1.4-1.1-3.0})\times 10^{-6},\\
Br(B^-\to f_0(980)K^{*-})=(15.4^{+1.6+1.4+4.1}_{-1.5-1.2-4.3})\times 10^{-6},
\en
where the uncertainties are from the decay constant of $f_0(980)$, the Gegenbauer moments $B_1$ and $B_3$. If $f_0(980)$
is purely composed of $n\bar n$, the branching ratios for $B\to f_0(980)K^*$ are:
\be
Br(\bar B^0\to f_0(980)\bar K^{*0})=(2.2^{+0.3+0.4+1.2}_{-0.2-0.5-1.1})\times 10^{-6},\\
Br(B^-\to f_0(980)K^{*-})=(4.0^{+0.5+0.5+1.6}_{-0.4-0.7-1.7})\times 10^{-6},
\en
where the uncertainties are from the same quantities as above.

The branching ratio for decay
$B^-\to f_0(980)K^{*-}$ in the upper extreme case is consistent with QCDF results \cite{hycheng2}:
\begin{eqnarray}
Br(B^-\to f_0(980)K^{*-})=\begin{cases}
14.3\times 10^{-6}, \text{for} f_0(980)=s\bar s,\\
6.9\times 10^{-6}, \text{for} f_0(980)=n\bar n.
\end{cases}
\end{eqnarray}

\begin{figure}[tb]
\begin{center}
\includegraphics[scale=0.65]{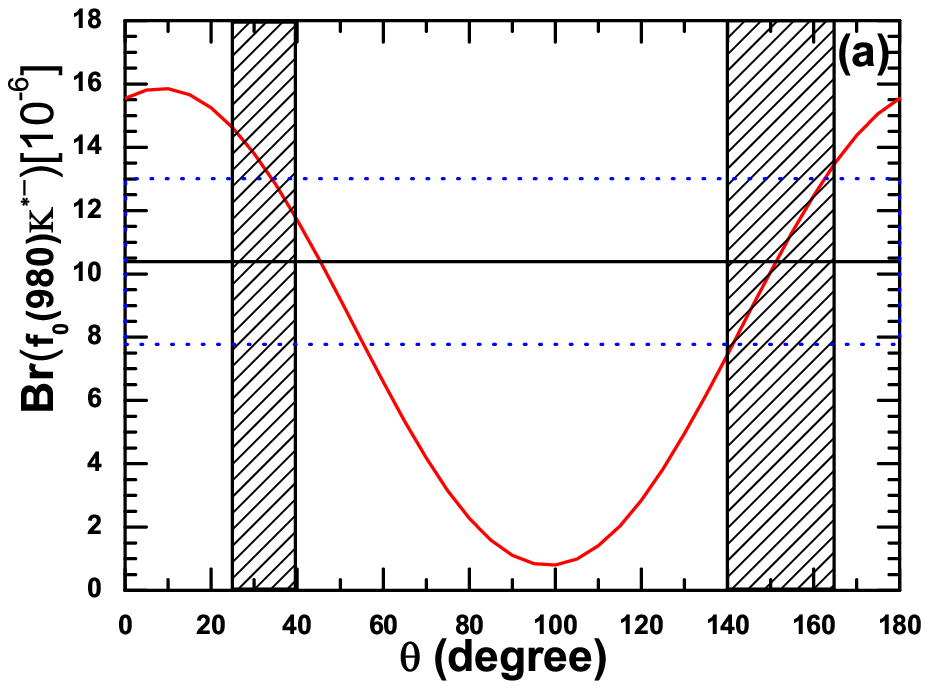}
\includegraphics[scale=0.65]{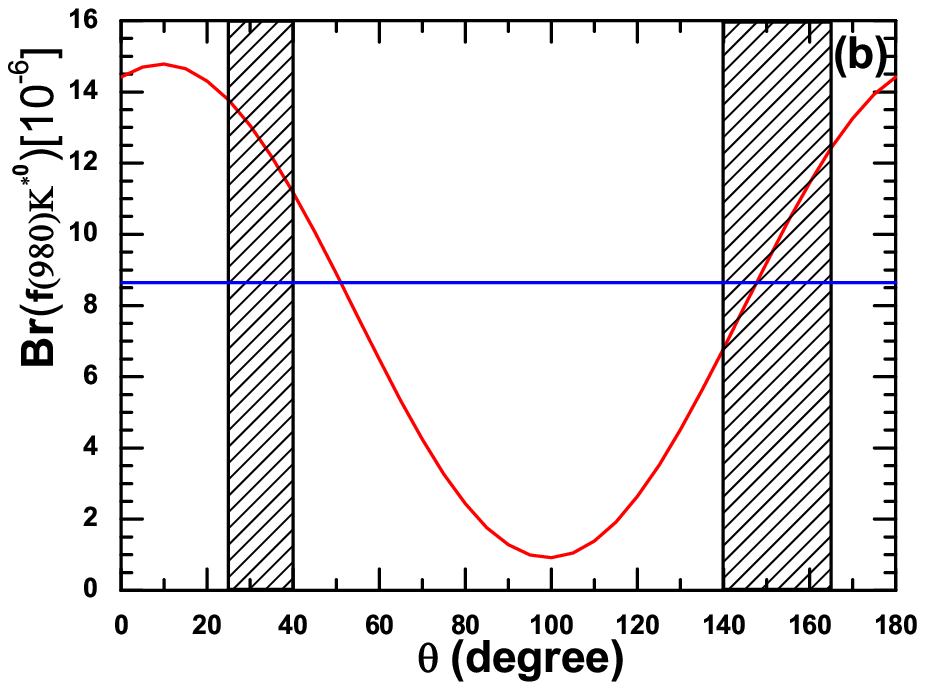}
\vspace{0.3cm}
\caption{The $\theta$ dependence of the branching
ratios (in units of $10^{-6}$) of the decays (a) $B^-\to f_0(980)K^{*-}$ and
(b) $\bar B^0\to f_0(980)\bar K^{*0}$. The horizontal solid lines show (a) the measured value and
(b) the experimental upper limit, respectively. The horizontal band within the doted
lines shows the experimentally allowed region of decay $B^-\to f_0(980)K^{*-}$. The vertical bands show
two possible ranges of $\theta$: $25^\circ<\theta<40^\circ$ and $140^\circ<\theta<165^\circ$.     }
\label{fig2}
\end{center}
\end{figure}

The Branching ratio of $B\to f_0(980)K^*$ depends on the mixing angle $\theta$ of strange and nonstrange components of the $f_0(980)$.
In Fig.\ref{fig2}, we plot the branching ratios as functions of the mixing angle $\theta$. Using the above mentioned range of the mixing angle, we
obtain:
\be
Br(B^-\to f_0(980)K^{*-})=(11.7\sim14.6)\times 10^{-6},\\
Br(\bar B^0\to f_0(980)\bar K^{*0})=(11.2\sim13.7)\times 10^{-6},
\en for $25^\circ<\theta<40^\circ$; as for the other range
$140^\circ<\theta<165^\circ$, these two branching ratios are: \be
Br(B^-\to f_0(980)K^{*-})=(7.5\sim13.5)\times 10^{-6},\\
Br(\bar B^0\to f_0(980)\bar K^{*0})=(6.7\sim12.5)\times 10^{-6}, \en
where only the central values of other input parameters are used. It is easy to see that the pQCD predictions can account
for the measured value or the experimental upper limit in the range $140^\circ<\theta<165^\circ$ (shown in Fig.2).
From the Fig.2(b), one can find the branching ratio of $\bar B^0\to f_0(980)\bar K^{*0}$
should be not far away from the upper limit (i.e. $8.6\times10^{-6}$). If we take $\theta=140^\circ$, the value of
$Br(\bar B^0\to f_0(980)\bar K^{*0})$ is about $6.8\times10^{-6}$,
which is consistent with the experimental value, $(5.2\pm 2.2)\times10^{-6}$ \cite{aubert}.
But for $25^\circ<\theta<40^\circ$, the predicted $\bar B^0\to f_0(980)\bar K^{*0}$ rate exceeds  the current
experimental limit.

\begin{table}
\caption{ Decay amplitudes for $\bar B^0\to f_0(980)\bar K^{*0}$
($\times 10^{-2} \mbox {GeV}^3$), where "this work" denotes the results using the distribution
amplitudes $\Phi_{f_0}$, $\Phi^S_{f_0}$ and $\Phi^T_{f_0}$ given in the previous section, "\cite{Chenf0K1}" denotes
the results using the DAs proposed in \cite{Chenf0K1}.}
\begin{center}
\begin{tabular}{cc|c|c|c|c}
\hline \hline  $\bar ss$&&$F^{f_0}_{e\bar K^{*0}}$ & $M^{f_0}_{e\bar K^{*0}}$ & $M^{f_0}_{a\bar K^{*0}}$ & $F^{f_0}_{a\bar K^{*0}}$ \\
\hline
This work &   &6.02&$1.12+4.37i$&$-0.45-0.78i$&$0.32+7.32i$ \\
\cite{Chenf0K1} &   &3.25&$0.29+0.31i$&$0.56-0.70i$&$-7.49+0.42i$ \\
\hline \hline
  $\bar nn $&&$F^{\bar K^{*0}}_{ef_0}$&$M^{f_0,T}_{e\bar K^{*0}}$ &  $M^{f_0}_{e\bar K^{*0}}$& $M^{\bar K^{*0}}_{ef_0}$ \\
\hline
This work &   &-6.74&$-19.47-59.17i$&$2.9+11.3i$&$0.81-0.56i$ \\
\cite{Chenf0K1} &   &10.5&$7.54+6.97i$&$0.27+0.28i$&$-0.37+1.81i$ \\
\hline\hline
  $\bar nn $&& $M^{\bar K^{*0}}_{af_0}$ & $F^{\bar K^{*0}}_{af_0}$ &&  \\
\hline
This work &   &$0.17+0.13i$&$0.35-6.77i$&--&--\\
\cite{Chenf0K1} &    &$0.14-0.07i$&$-7.42+0.19i$ &--&--\\
\hline\hline
\end{tabular}\label{amp}
\end{center}
\end{table}

Our results are larger than the previous pQCD results \cite{Chenf0K1}. Part of the reason is in taking the different parameters, for example the decay constant
of $f_0(980)$. The main reason is that the author in \cite{Chenf0K1} neglected the twist-2 contribution but only used the twist-3 distribution amplitude
$\phi_f^S(x)$, which is symmetry for $x\to 1-x$. Taking these shapes of distribution amplitude would make the contributions from the non-factorizable
diagrams (c) and (d) cancel with each other. But here we include the twist-2 distribution amplitude and use
the asymptotic form of the twist-3 distribution amplitude. In this
case, the contributions from $f_0$ emission non-factorizable diagrams are large. In order to
show this character, we list the numerical results for different topology diagrams of $\bar B^0\to f_0(980)K^{*0}$
in Table II. In the table, $F^{f_0}_{e(a)\bar K^{*0}}$ and $M^{f_0}_{e(a)\bar K^{*0}}$ denote as
the contributions from $f_0$ emission (annihilation) factorizable contributions and non-factorizable contributions
from penguin operators respectively. Similarly, $F^{\bar K^{*0}}_{e(a)f_0}$ and $M^{\bar K^{*0}}_{e(a)f_0}$ are the
$\bar K^{*0}$ emission (annihilation) factorizable contributions and non-factorizable contributions from penguin
operators, respectively. $M^{f_0,T}_{e(a)\bar K^{*0}}$ denote the $f_0$ emission non-factorizable contribution from
tree operator $O_2$.
It is easy to see that $M^{f_0}_{e(a)\bar K^{*0}}$ and $M^{f_0,T}_{e(a)\bar K^{*0}}$ obtain
an enhancement compared to previous estimates.
It suggests that the non-factorizable type amplitude is sensitive
to the shape of the distribution amplitudes.

%=================================================================

Now we turn to the evaluations of the direct CP-violating asymmetries of $B^-\to f_0(980)K^{*-}$ and
$\bar B^0\to f_0(980)\bar K^{*0}$ decays in the pQCD approach. The direct CP-violating asymmetry can
be defined as
\be
\acp^{dir}=\frac{ |\overline{\cal M}|^2-|{\cal M}|^2 }{
 |{\cal M}|^2+|\overline{\cal M}|^2}\;.
 \en
%%%%%%%%%%%%%%%%%%%%%%%%%%%%%%%%%%%%%%%%
\begin{figure}[tb]
\begin{center}
\includegraphics[scale=0.65]{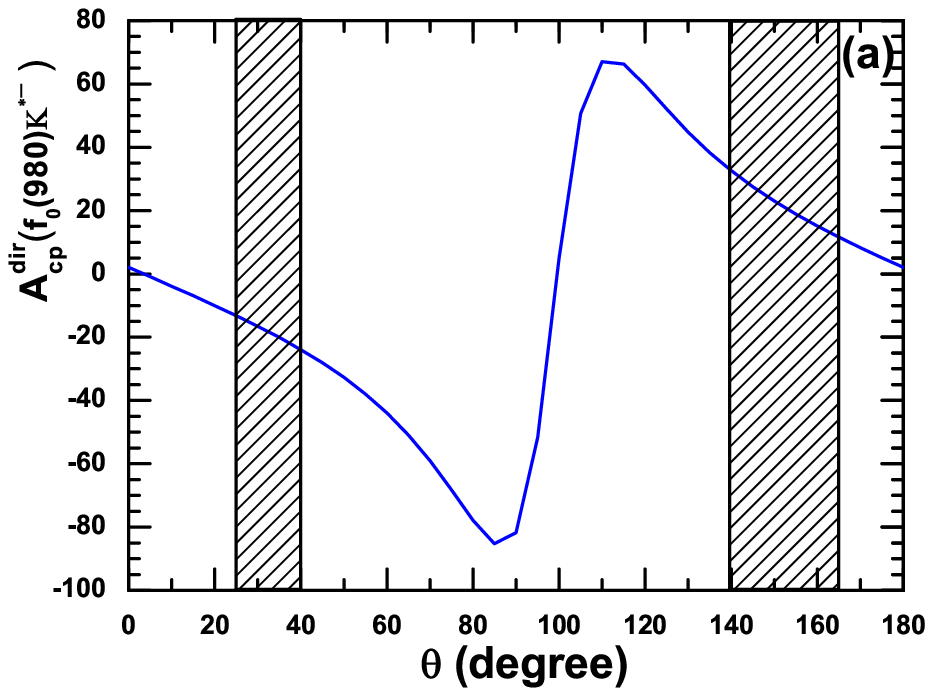}
\includegraphics[scale=0.65]{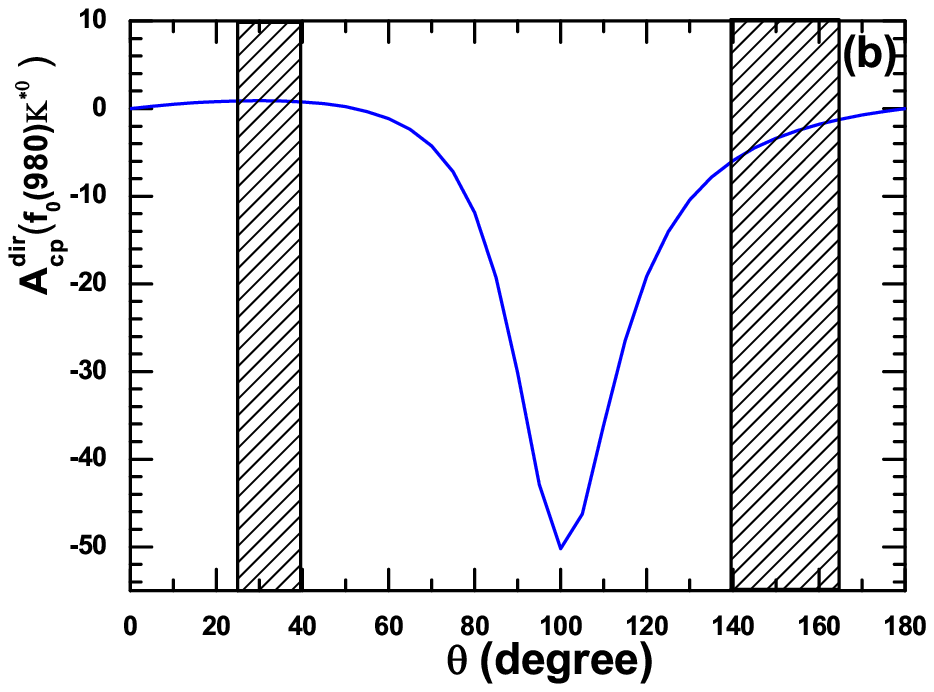}
\vspace{0.3cm}
\caption{The $\theta$ dependence of the direct CP asymmetry
(in units of percent) of  the decays (a) $B^-\to f_0(980)K^{*-}$ and
(b) $\bar B^0\to f_0(980)\bar K^{*0}$. The vertical bands show
possible ranges of $\theta$: $25^\circ<\theta<40^\circ$ and $140^\circ<\theta<165^\circ$.}
\label{fig2}
\end{center}
\end{figure}
For the decay $\bar B^0\to f_0(s\bar s)\bar K^{*0}$, there is no tree contribution at the leading order,
so the CP asymmetry is naturally zero. But the
CP asymmetry of $\bar B^0\to f_0(n\bar n)\bar K^{*0}$ is large,
for the $f_0$ emission non-factorizable diagrams (Fig.1(c) and (d)) give the large tree
contributions, and its the direct CP asymmetry is about $-39\%$. It is similar to the decay
$B^-\to f_0(980)K^{*-}$. From the Fig.3(a), one can find that if taking the
mixing angle $25^\circ<\theta<40^\circ$, the direct CP asymmetry of the decay $B^-\to f_0(980)K^{*-}$ is:
\be
{\cal A}^{dir}_{CP}(B^-\to f_0(980)K^{*-})=(-15\sim-25)\%,
\en
which may suffice to explain the experimental result \cite{pdg08}:
\be
{\cal A}^{dir}_{CP}(B^-\to f_0(980)K^{*-})=(-34\pm21)\%.
\en
But if we take the mixing angle $140^\circ<\theta<165^\circ$, the value has the opposite sign with
the experimental result and becomes $(10\sim33)\%$. Certainly, the errors from both the experimental result and
the prediction are large.

From the upper analysis to the branch ratios of the decays $\bar B^0\to f_0(980)\bar K^{*0}$, $B^-\to f_0(980)K^{*-}$, it supports the conclusion
that the mixing angle should be in the range of $140^\circ<\theta<165^\circ$. But unfortunately the range of
$25^\circ<\theta<40^\circ$ as it seems cannot be ruled out absolutely. From Fig.2 and Fig.3, one can find
there exist some symmetries for these two angle ranges. Within (large) theoretical errors,
the results for the two angle ranges are both in agreement with the data. For example, if we
take the angle $25^\circ<\theta<40^\circ$ in the Fig.3(b), the direct CP asymmetry of the decay $B^-\to f_0(980)K^{*-}$ is:
\be
{\cal A}^{dir}_{CP}(B^-\to f_0(980)K^{*-})=(0.8\sim0.9)\%,
\en
and ${\cal A}^{dir}_{CP}(B^-\to f_0(980)K^{*-})=(-1.2\sim-5.9)\%$ for $140^\circ<\theta<165^\circ$. That is to say the values
of ${\cal A}^{dir}_{CP}(B^-\to f_0(980)K^{*-})$ for these two $\theta$ angle ranges
are close and both small.
%===========================================================================
%                 Conclusion
%============================================================================

\section{Conclusion}\label{summary}

In this paper, we calculate the branching ratios and CP-violating
asymmetries of $ \bar{B}^0\to f_0(980)\bar K^{*0}$ and $B^-\to f_0(980)K^{*-}$ decays
in the pQCD factorization approach by identifying $f_0(980)$ as the composition of $s\bar s$ and
$n\bar n=(u\bar u+d\bar d)/\sqrt 2$. Using the decay constants and light-cone distribution amplitude
derived from the QCD sum-rule method, we find that:
\begin{itemize}
\item
After including the twist-2 distribution amplitude and using the asymptotic form of twist-3
distribution amplitude, our results are larger than the previous pQCD predictions and can explain
the present experimental data or the upper limit.

\item
From the results, it indicates that the contributions from the non-factorizable $f_0$ emission type diagrams are large,
at the same time this type amplitude is sensitive to the shape of the distribution amplitudes.

\item
The branching ratio of $B\to f_0(980)K^*$ depends on the mixing angle $\theta$ of strange and nonstrange
components of the $f_0(980)$. One can find that there exit some symmetries for the values in the two angle ranges
(i.e., $25^\circ<\theta<40^\circ$
and $140^\circ<\theta<165^\circ$). So it is difficult to confirm the value of the mixing angle, unless we
can get enough and precise experimental data.
\item
For the neutral decay $\bar{B}^0\to f_0(980)\bar K^{*0}$, we predict that the direct CP-violating asymmetry is small,
only a few percent, which
can be tested by the future B factory experiments.
\end{itemize}

\section*{Acknowledgment}
This work is partly supported by Foundation of Henan University of Technology under Grant No.150374. Z.Q.~Zhang
would like to thank Wei Wang for reading the manuscript and for helpful discussions.
%%%%%%%%%%%%%%%%%%%%%%%%%%%%%%%%%%%%%%%%%%%%%%%%%%%%%%%%%%%%%%%%%%%%%%%%
%                               references
%%%%%%%%%%%%%%%%%%%%%%%%%%%%%%%%%%%%%%%%%%%%%%%%%%%%%%%%%%%%%%%%%%%%%%%%

\end{document}